\documentclass[aps,prl,preprint,floatfix]{revtex4-1}
\usepackage{amsmath}
\usepackage{amsfonts}
\usepackage{amssymb}
\usepackage{bm}
\def\*#1{\mathbf{#1}}
\usepackage[usenames,dvipsnames]{color} 
\usepackage{ulem}
\usepackage{graphicx}
\usepackage{epstopdf}

\usepackage{csquotes}
\usepackage[sort&compress]{natbib}

\usepackage [english]{babel}
\usepackage[utf8]{inputenc}
\usepackage[T1]{fontenc}

\usepackage{gensymb}

\usepackage{lineno}

\usepackage{hyperref}
\hypersetup{
  colorlinks   = true, 
  urlcolor     = black, 
  linkcolor    = red, 
  citecolor   = blue 
}

\usepackage{tabularx}
\newcolumntype{Y}{>{\centering\arraybackslash}X}  

\begin{document}

\title{Roughness and scaling properties of oxide glass surfaces at the nanoscale
}

\author{Zhen Zhang}
\affiliation{Laboratoire Charles Coulomb (L2C), 
University of Montpellier and CNRS, F-34095 Montpellier, France}

\author{Simona Ispas}
\affiliation{Laboratoire Charles Coulomb (L2C), 
University of Montpellier and CNRS, F-34095 Montpellier, France}

\author{Walter Kob}
\email[Corresponding author: ]{walter.kob@umontpellier.fr}
\affiliation{Laboratoire Charles Coulomb (L2C),
University of Montpellier and CNRS, F-34095 Montpellier, France}

\date{\today}

\begin{abstract}  
Using atomistic computer simulations we determine the roughness and
topographical features of melt-formed (MS) and fracture surfaces (FS)
of oxide glasses. We find that the topography of the MS is described
well by the frozen capillary wave theory. The FS are significant rougher
than the MS and depend strongly on glass composition. The height-height
correlation function for the FS shows an unexpected logarithmic dependence
on distance, in contrast to the power-law found in experiments. We thus conclude that on length scales less than 10~nm FS are not self-affine fractals.

\end{abstract}


\maketitle

Surface roughness plays a crucial role for the
functional properties of a material, including
friction~\cite{persson_nature_2005,urbakh2004nonlinear},
adhesion~\cite{pastewka2014contact} and transport
properties~\cite{gotsmann2013quantized}. Understanding the
nature and modifying this roughness is thus of great practical
importance. In comparison with the surfaces of crystalline materials, the surfaces of amorphous materials such as glasses has received much less attention since the
disorder renders the probing and characterization of such systems
difficult~\cite{hench1978physical,Pantano1989,bach1997advanced,bocko1991surface,dey2016cleaning,zheng2019protein}.
Since a glass is an out of equilibrium system, the properties of its
surface depends on the process by which it has been produced. Usually
one considers two types of pristine (i.e., without post-processing)
glass surfaces: i) Melt-formed surfaces (MS) which result from cooling
a liquid with a free surface to the solid state and ii) Fracture surfaces
(FS) resulting from a mechanical failure.

The topography of a MS is often described using the concept of a frozen
liquid interface~\cite{jackle_intrinsic_1995,seydel_freezing_2002},
i.e.,~upon cooling the sample, the capillary waves at the surface freeze at a
temperature $T_0$. Thus the roughness of a pristine MS is predicted to
be $\sigma=\sqrt{ k_{\rm B}T_0/\gamma}$, where $\sigma$ is the standard
deviation of the surface height fluctuation, $k_{\rm B}$ is Boltzmann's
constant, and $\gamma$ is the surface tension at $T_0$. Atomic force
microscope (AFM) experiments on oxide glass surfaces have shown that
this prediction works well if one uses for $T_0$ the glass transition
temperature~\cite{sarlat_frozen_2006,gupta_nanoscale_2000,roberts2005ultimate}.
This theoretical framework also predicts that the height-height
correlation function

\begin{equation}
\Delta z(r) = \sqrt{\big \langle [z(r+x) - z(r)]^2 \big \rangle_x} \quad,
\label{eq1}
\end{equation}

\noindent
which gives the height difference between two points separated by a distance $r$
along a direction $x$, increases like $(\Delta z)^2 \propto$ ln~$r$. This
logarithmic dependence was validated experimentally with $r$
ranging from around 10~nm to 1000~nm~\cite{sarlat_frozen_2006}.  However,
for $r<10$~nm the dependence on $r$ is basically unknown.

Describing the topography of the FS is more
difficult than that for the MS since the former results from a highly non-linear process
which involves a complex interplay between heterogeneities in the composition,
microstructure, and mechanical properties present in a glass (see,
e.g.~\cite{bouchaud_scaling_1997,ciccotti_stress-corrosion_2009,bonamy2011failure}
for reviews).  Experimental studies of oxide glasses
have shown that the roughness of the FS depends strongly on the
composition~\cite{gupta_nanoscale_2000,wiederhorn_roughness_2007,pallares_roughness_2018}
and is larger than the one found in MS~\cite{gupta_nanoscale_2000}. AFM
measurements have given evidence that the FS of various materials can
be described as self-affine fractals~\cite{mandelbrot_fractal_1984},
i.e.,~the height-height correlation function scales like $\Delta z
\propto r^{\zeta}$. Here $\zeta$ is the roughness (or Hurst) exponent
which was found to depend on the fracture mode, the length scale considered, and the
material~\cite{milman_fracture_1994,bonamy_scaling_2006,ponson_two-dimensional_2006,pallares_roughness_2018,wiederhorn_roughness_2007}.
However, whether this self-affine description for the FS holds also
down to the nanometer scale is still an open question since at such small
scales the finite size of the probing tip severely restricts the spatial
resolution~\cite{lechenault_effects_2010,mazeran2005curvature,schmittbuhl_reliability_1995}.

Here, using large scale atomistic simulations, we
provide quantitative insight into the topographical features of oxide
glasses for the length scales ranging from a few Angstr\"{o}ms to several
tens of nanometers. In particular we analyze the morphology, roughness,
symmetry, and statistical scaling of the MS and FS, i.e.,~surfaces which
have very distinct manufacturing histories.

We investigate two archetypical compositions for oxide glasses, namely
pure SiO$_2$ and binary Na$_2$O-$x$SiO$_2$, with $x = 3, 4, 5, 7, 10,$
and 20. The atomic interactions are described by a two-body effective
potential~(SHIK)~\cite{sundararaman_new_2018,sundararaman_new_2019}
which has been shown to give a reliable description of the
structural, mechanical and surface properties of sodium silicate
glasses~\cite{zhang_potential_2020,zhang_surf-vib_2020}. Our samples
contain typically $2.3\times10^6$ atoms, corresponding to box sizes
of around 20~nm$\times$30~nm$\times$50~nm (in the $x$, $y$, and $z$
directions, respectively). Periodic boundary conditions were applied in
the $x$ and $y$ directions while in the $z$-direction we introduced (in
the melt) two free surfaces. The samples were melted and equilibrated at
a high temperature (composition dependent) and then cooled down to 300~K
under zero pressure. In the following we will refer to these two surfaces,
generated by the melt-quench procedure, as MS. The glass samples were
then subjected to an uniaxial tension with a strain rate of 0.5~ns$^{-1}$
in the $y$-direction until complete fracture occurred, creating thus
two FS. The surface atoms were identified by using a well-established
geometric method~\cite{edelsbrunner_three-dimensional_1994} which allows
thus to study the topographical features of the surfaces. More details
on the sample preparation and surface construction are given in the
Supplemental Material (SM).

Figure~\ref{fig1_surf-morphology} shows the (color coded) height
fluctuations of the surfaces for three representatives compositions:
Silica, NS10 ($\approx$9\% Na$_2$O), and NS3 (25\% Na$_2$O). The top and
bottom panels are for the MS and FS, respectively and the $z=0$ level
has been determined such that the mean of the fluctuations is zero.
For the MS, panels (a-c), one recognizes that the amplitude of these
height fluctuations seems to be independent of the composition, and
that also the spatial extent of the structures are independent of the
Na$_2$O content. In contrast to this, the FS, panels (d-f), show height fluctuations
that are larger than the ones for the MS and clearly grow in amplitude
and extent if the concentration of Na$_2$O increases.
Also one recognizes that the surfaces seem to be anisotropic and in the
following we will quantify these observations.


\begin{figure}[t]
\includegraphics[width=\columnwidth]{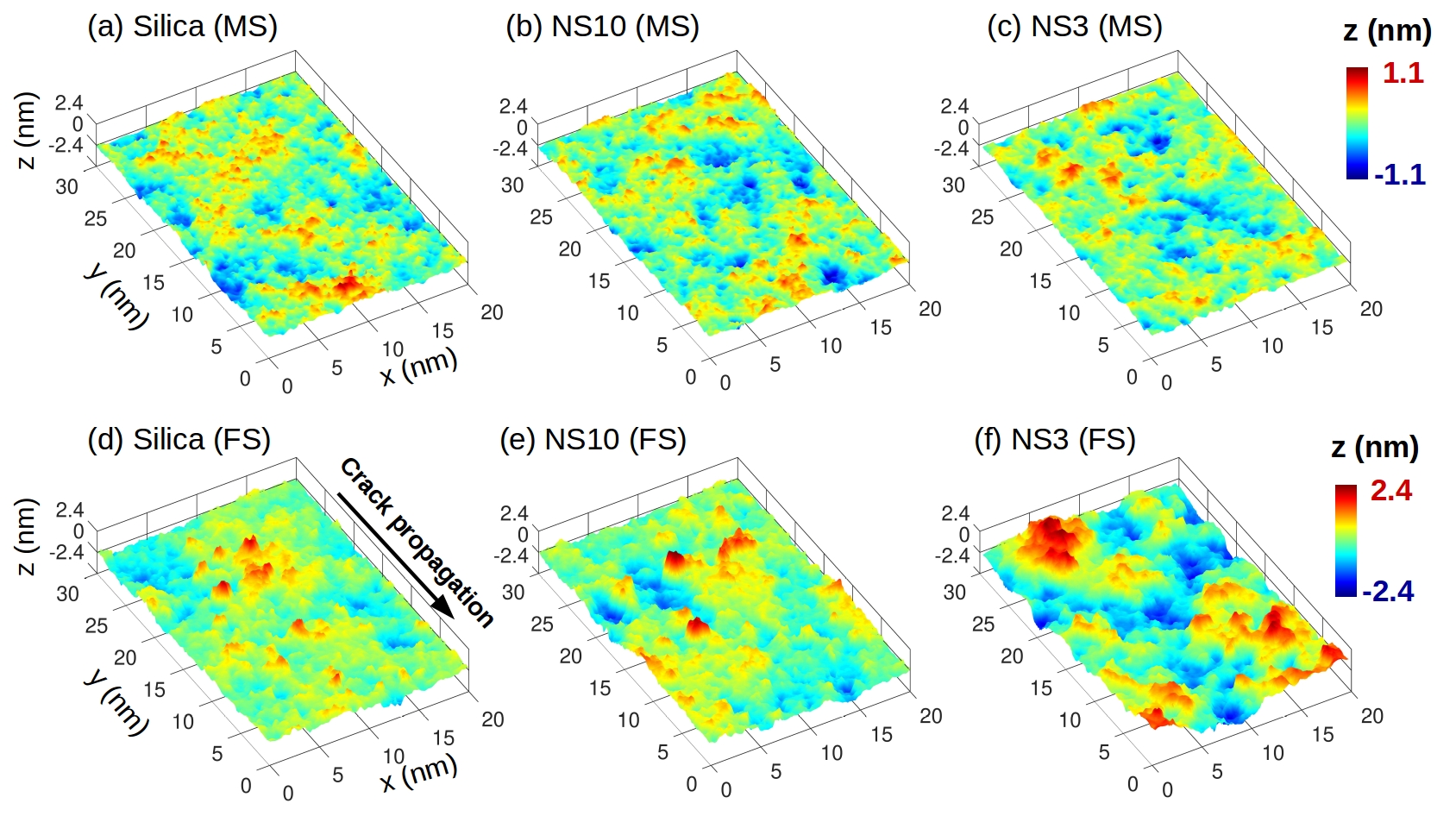}
\caption{Surface morphology. Melt-formed surfaces (a-c) and fracture surfaces (d-f) for three representative glass compositions. For the FS, the crack propagates in the negative $y$-direction and the crack front is parallel to the $x$-direction. From left to right the compositions are silica, NS10, and NS3, respectively.
}
\label{fig1_surf-morphology}
\end{figure}

\begin{figure}[ht]
\includegraphics[width=0.95\columnwidth]{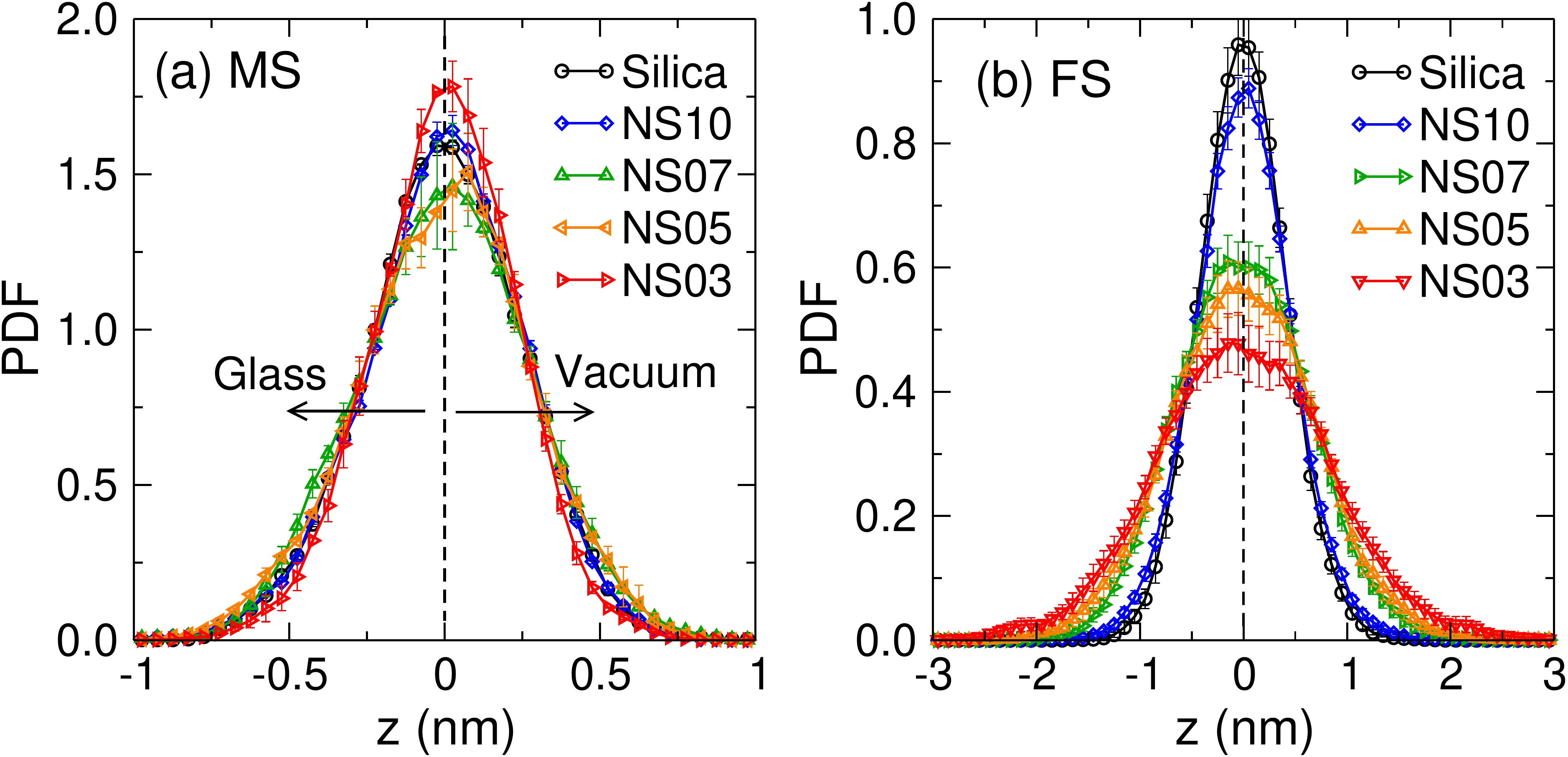}
\includegraphics[width=0.95\columnwidth]{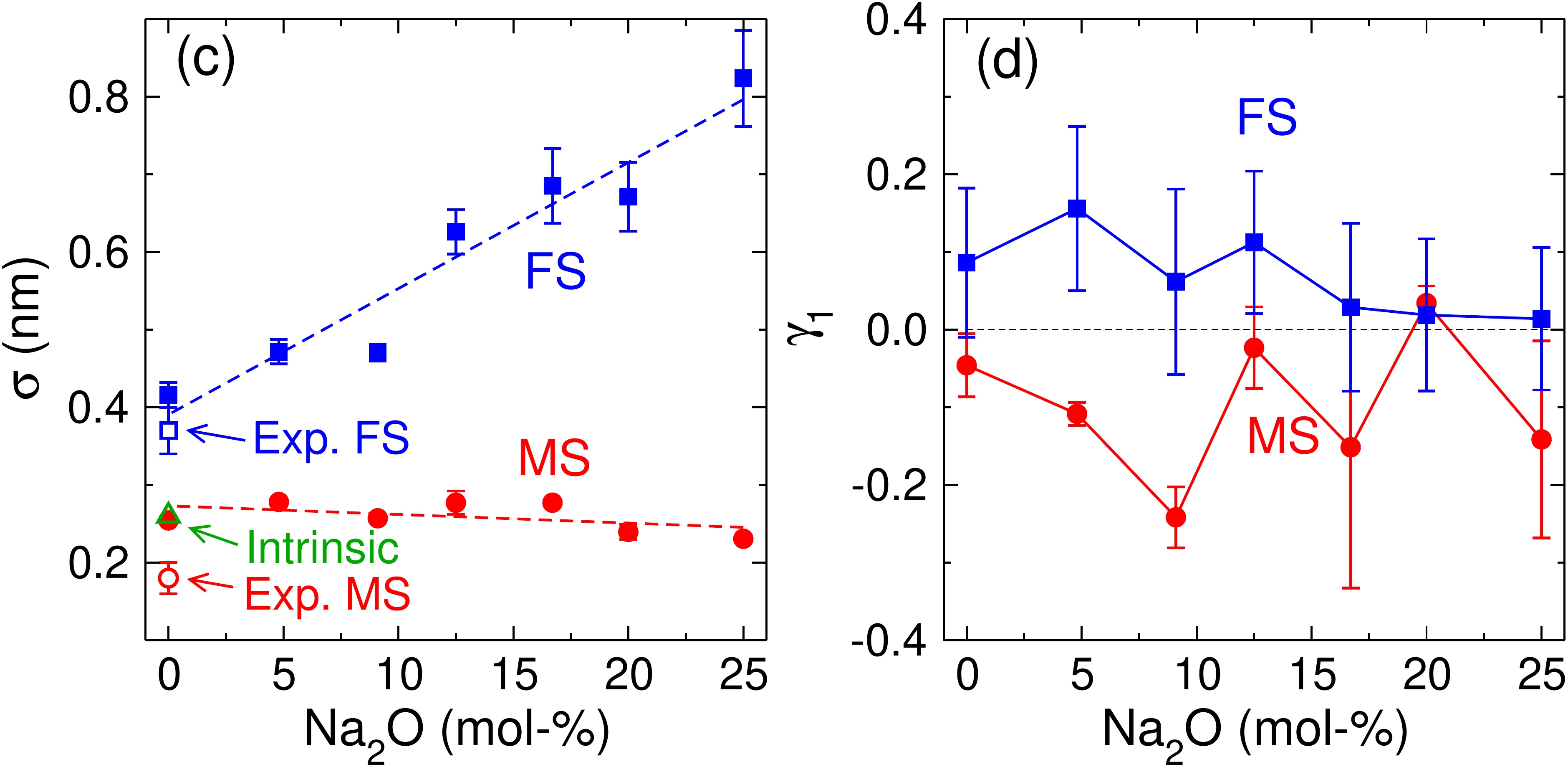}
\caption{(a) and (b): Surface height distribution as a function of
the composition. The mean surface height $\big \langle z \big \rangle$
is equal to zero. Heights with $z>0$ point towards the vacuum.
(c) Roughness of the
surfaces. The dashed lines are linear fits to the data. Experimental
data are from Ref.~\cite{gupta_nanoscale_2000}. 
The triangle corresponds to the intrinsic roughness of silica surface as estimated from the theory of frozen capillary waves. (d) Skewness of the surface height distribution.}
\label{fig2_surf-height-distri}
\end{figure}

Figures~\ref{fig2_surf-height-distri}(a) and (b) show the distribution
of the surface height $z$ for different glass compositions. For
the MS, panel (a), we find that this distribution is basically
independent of the composition, in agreement with the snapshots
shown in Fig.~\ref{fig1_surf-morphology}. In contrast to this, the
distribution for the FS shows a clear dependence on the composition
in that it becomes wider with increasing Na$_2$O concentration
(i.e., smaller $x$). The change of the surface height distribution
is directly related to the surface roughness $\sigma$, which is
defined as the standard deviation of height fluctuations. (Note that
$\sigma$ of the FS is basically independent of the distance from the
fracture origin, see Fig.~\ref{SI_fig1_surface-segment-roughness}.)
Figure~\ref{fig2_surf-height-distri}(c) shows $\sigma$ as a function
of the mole concentration of Na$_2$O. For the MS, $\sigma$ is around
0.25~nm for silica and 0.23~nm for NS3, thus showing indeed a very mild
dependence on the composition.  This observation is likely related
to the fact that the MS are rich in Na~\cite{zhang_surf-vib_2020},
i.e.~a species that plastifies the glass, thus allowing to smooth out
even small irregularities.  Using the capillary wave theory mentioned
above it is possible to estimate the roughness
of the MS from the surface tension, and for the case of silica one finds $\sigma\approx0.26$~nm
(data point labelled ``intrinsic'')~\cite{gupta_nanoscale_2000}, in very
good agreement with our simulation result.

The roughness of the FS is higher than the one for the MS, in agreement
with the qualitative impression given by Fig.~\ref{fig1_surf-morphology},
and shows a clear dependence on the Na$_2$O content in that it
increases from $\approx0.42$~nm for silica to $\approx0.82$~nm
for NS3.  The increase of $\sigma$ with Na$_2$O concentration
can be rationalized by the fact that, with the addition of Na,
the glass becomes increasing ductile when subjected to mechanical
loading~\cite{wang_intrinsic_2015,pedone_dynamics_2015,zhang_potential_2020}.
This increase of ductility originates from the enhanced
heterogeneities in the micro-structure and the local
plasticity of the glass, leading to a rougher fracture
surface~\cite{wang_nanoductility_2016,zhang_thesis_2020}. (See
Fig.~\ref{SI_fig2_ns0-FS-srate-effect}(a) for the dependence of $\sigma$
on the strain rate.)

Also included in the graph are the experimental values of
the roughness for silica glass surfaces as obtained from AFM
measurements~\cite{gupta_nanoscale_2000}. One observes that these
experimental data are somewhat below our simulation values and the
theoretical prediction (for the MS). This discrepancy might be due
to the insufficient spatial resolution of this experimental technique
(see also the discussion below).

A further property of interest is the symmetry of the surfaces, which
can be quantified by the skewness $\gamma_1$ of the surface height
distribution. The question of interest is whether or not the two sides
of the surface (facing the vacuum/facing the glass) are statistically
equivalent.  Figure~\ref{fig2_surf-height-distri}(d) shows that the MS
have a negative $\gamma_1$, i.e.~there are more deep holes than high
protrusions, while for the FS $\gamma_1$ is positive, i.e.~there
are more high protrusions than deep holes. The result for the FS is
coherent with the view that during the fracture process the breaking
of bridges or chain-like structures gives rise to a spiky surface. Note
that a non-vanishing $\gamma_1$ indicates that the capillary wave theory
cannot be strictly valid since this approach predicts $\gamma_1=0$. (See
Fig.~\ref{SI_fig3_nsx-surf-roughness-Tdepend} for the surface properties
at elevated temperatures.)

\begin{figure}[bt]
\centering
\includegraphics[width=0.95\columnwidth]{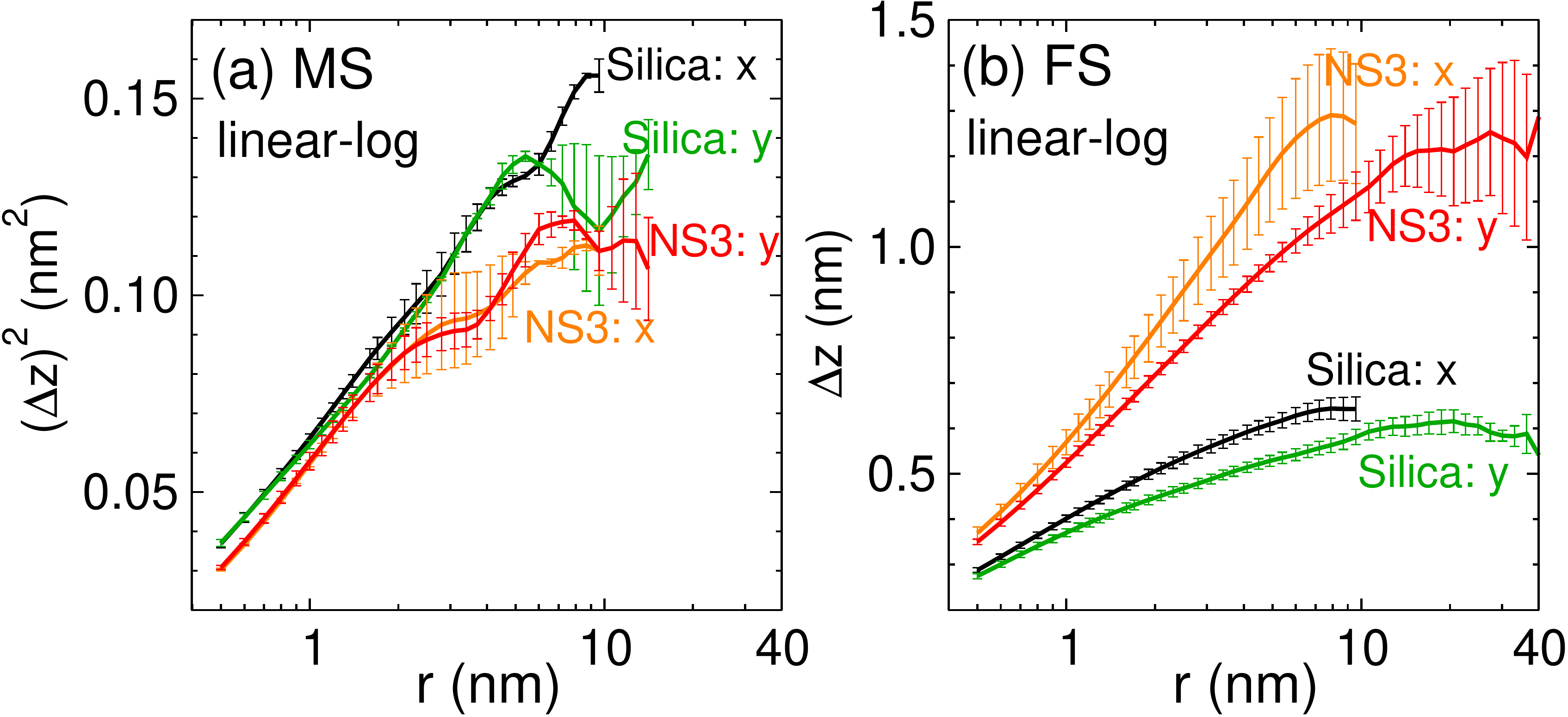}
\includegraphics[width=0.95\columnwidth]{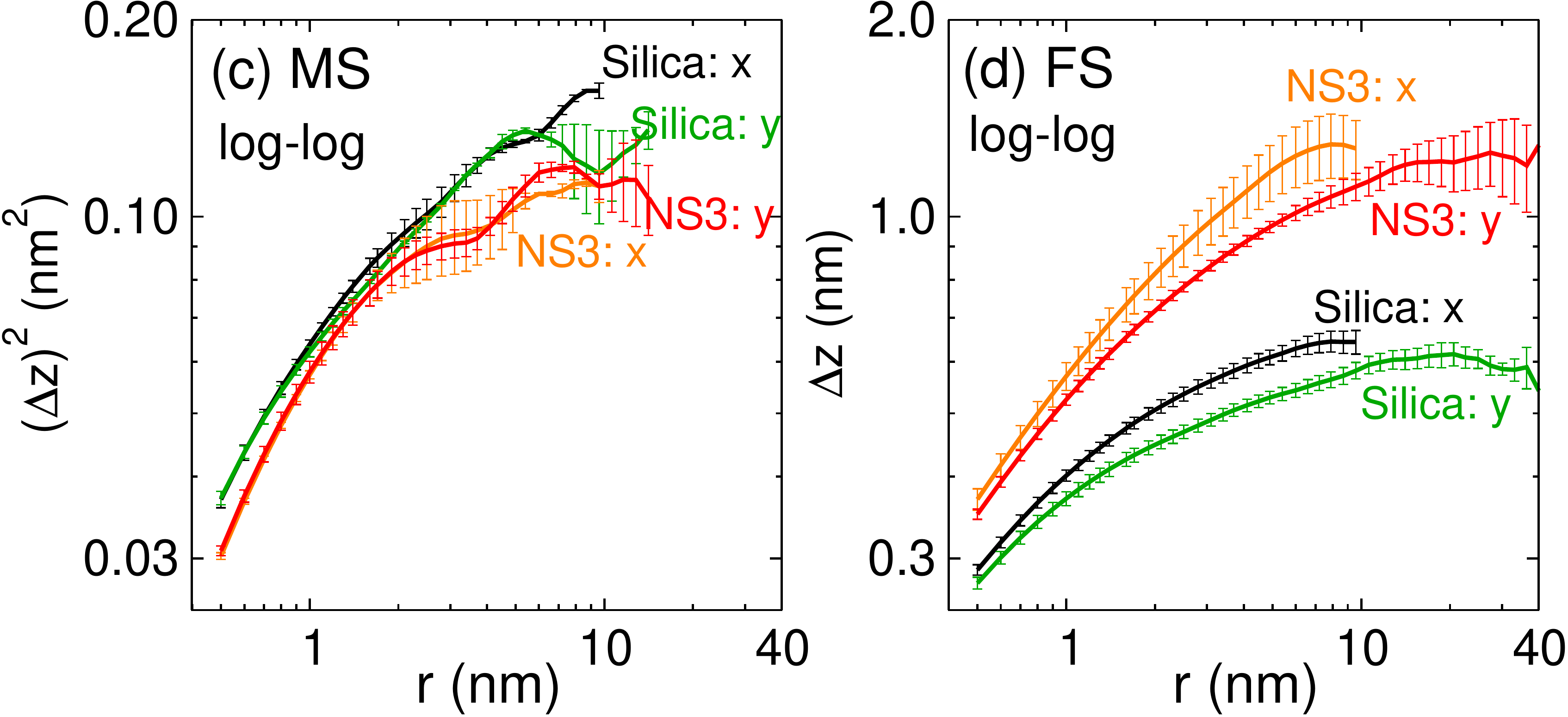}
\caption{(a) and (b): Surface height correlation function (linear-log
scale) for the MS (a) and FS (b).  (c) and (d): Same data as in
(a) and (b) but on log-log scale. Note that the ordinates for the left
and right panels are not the same. The labels $x$ and $y$ correspond
to the direction parallel to the crack front and the direction of crack
growth, respectively.}
\label{fig3_surface-height-corr}
\end{figure}

To characterize the structure of the surfaces on larger scales it is
useful to look at the height-height correlation function defined in
Eq.~(\ref{eq1}). Figure~\ref{fig3_surface-height-corr}(a) shows $(\Delta
z)^2$ as a function of $r$ for the MS of silica and NS3.  Two (orthogonal)
directions are considered and, as expected, they give the same result,
indicating that the MS is isotropic. Moreover, we note that the curves
for silica are slightly above the ones for NS3, indicating that the
MS of silica is a bit rougher than the one of NS3, in agreement with
Fig.~\ref{fig2_surf-height-distri}(c). One also observes that $(\Delta
z)^2$ increases logarithmically with $r$, in agreement with the prediction
of the frozen capillary wave approach~\cite{jackle_intrinsic_1995}. Since
AFM experiments on MS have found the same $r$-dependence for $r\geq
10$~nm~\cite{sarlat_frozen_2006}, we can conclude that this theory gives
a reliable description for length scales that range from the atomic
(sub-nanometer) to the micrometer scale.

For the FS, Fig.~\ref{fig3_surface-height-corr}(b), we find that
$\Delta z(r)$ (no square!) shows a linear increase with $\ln r$,
thus a dependence that is very different from the one found for
the MS. In this case the roughness depends on the direction in
which it is measured. The curve for the $x$-direction (parallel
to the crack front) is about 15\% higher than the one for the
$y$-direction (orthogonal to the crack front), irrespective of glass
composition. In addition, the slope of the linear regime depends
not only on the composition but also on the probing direction. This
results indicate that the FS is anisotropic, and its roughness depends
on the composition, in agreement the experimental findings on the FS of oxide
glasses~\cite{ponson_two-dimensional_2006,pallares_roughness_2018,wiederhorn_roughness_2007}.
Note that at large $r$ all of the curves tend to saturate, a
behavior that is most likely related to the fact that the sample
is finite and hence fluctuations are bounded. The parameters
obtained from a logarithmic fit to the small $r$ data in panels
(a) and (b), as well as for other compositions, are shown in
Fig.~\ref{SI_fig4_nsx-surf-hh-MS-FS-fitting}. We also confirmed that the
strain rate used for the fracture simulation has only a weak effect on
these parameters [Fig.~\ref{SI_fig2_ns0-FS-srate-effect}(b)].

In Fig.~\ref{fig3_surface-height-corr}(c) and (d) we show on log-log
scale the same data as plotted in Fig.~\ref{fig3_surface-height-corr}(a)
and (b), respectively. It is evident that this type of presentation
of the data does not rectify it, demonstrating that on the length
scales we have explored the height-height correlation is not given
by a power-law, i.e.~neither surface has the characteristics of a
fractal. Instead, the $\ln r$-dependence we find for the FS is
in qualitative agreement with theoretical and numerical studies on the
fracture surface of heterogeneous materials (mode I fracture, i.e.~tensile
loading)~\cite{ramanathan_quasistatic_1997,bares_nominally_2014}.

Our finding that the FS cannot be described as a self-affine
fractal on the length scales we have considered
is at odds with AFM measurements that have reported a
power-law dependence of $\Delta z(r)$ down to the scale of
1~nm~\cite{ponson_two-dimensional_2006,pallares_roughness_2018}. To
elucidate the origin of this discrepancy one has to recall that the size
of an AFM tip is finite which limits the lateral resolution of the
measurements~\cite{lechenault_effects_2010,mazeran2005curvature}
and can induces biases in the characterization of the
surface~\cite{schmittbuhl_reliability_1995}.

\begin{figure}[th]
\centering
\includegraphics[width=0.95\columnwidth]{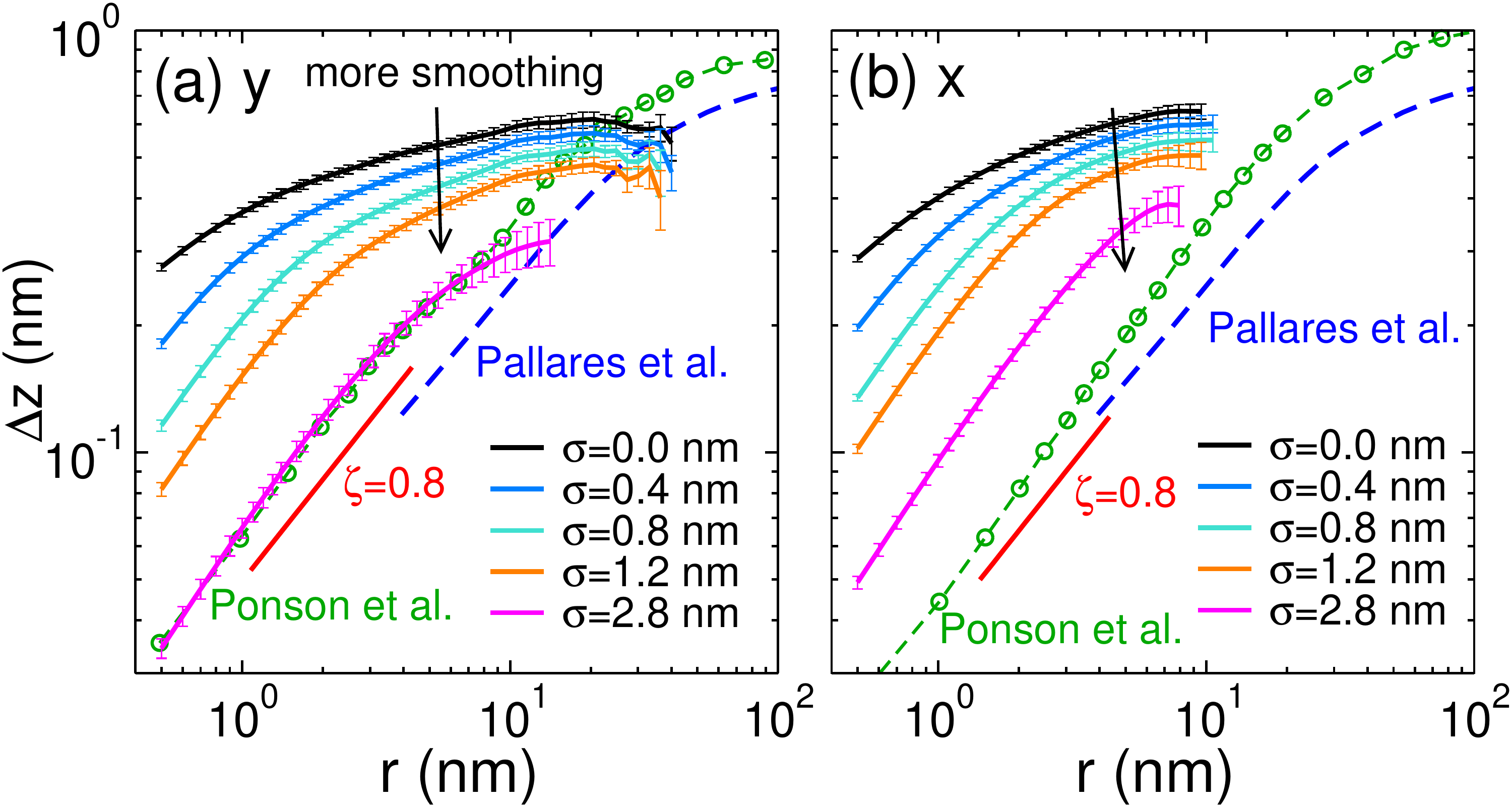}
\caption{Influence of spatial resolution on the surface height correlation
for the FS of SiO$_2$. (a) and (b) are for the direction orthogonal
and parallel to the crack front, respectively.  
A Gaussian filter (width $\omega$) was applied to smooth
the surface. The curves labeled $\omega=0$
correspond to the original data. Also included are experimental data
from AFM measurements of fracture surfaces produced by sub-critical crack
propagation~\cite{ponson_two-dimensional_2006,pallares_roughness_2018}.
}
\label{fig4_surface-smooth-effect}
\end{figure}

In order to investigate the effect of spatial resolution
we have convoluted our FS with a two-dimensional Gaussian
filter~\cite{pitas2000digital} of width $\omega$ (see SM) and
then recalculated the height-height correlation function for this
smoothed surface. In Fig.~\ref{fig4_surface-smooth-effect} we show
for the case of the FS of silica the resulting correlation functions
for different values of $\omega$. The curve $\omega=0$ corresponds
to the original (non-smoothed) data. We find that with increasing
$\omega$ the value of $\Delta z$ decreases significantly since the
convolution irons out the deep holes/high spikes. Surprisingly,
we note that at small $r$ the convoluted signal can be described
well with a power-law, and that the $r$-range in which this
functional form is observed increases with $\omega$ while the
exponent $\zeta$ is independent of $w$. For $\omega=2.8$~nm, the
correlation function of the convoluted surface in the $y$-direction
[Fig.~\ref{fig4_surface-smooth-effect}(a)] matches very well the AFM
data by Ponson~{\it et al.}~\cite{ponson_two-dimensional_2006}. The
exponent of the power-law is $\zeta\approx0.8$, i.e. the
claimed ``universal'' roughness exponent found in previous experimental
studies~\cite{ponson_two-dimensional_2006,pallares_roughness_2018,bonamy_scaling_2006,bouchaud_scaling_1997}.
The data in the $x-$direction, panel (b), shows qualitatively the same
variation as the ones in the $y-$direction. However, we find that one
needs to apply stronger smoothing to match quantitatively the convoluted
surfaces with the experimental ones, a result that is related to the
fact that the FS is anisotropic and the surface profile in the direction
parallel to the crack front is rougher than the one in the $y-$direction.
These results indicate that the power-law observed in experiments on the
scale of a few nanometers might be an artifact of insufficient resolution
of the surface measurements and that in reality the correlation function
takes higher values than that extracted from AFM studies.

In conclusion, the results of this work reveal how the topographical
features of glass surfaces reflect the way they have been produced. While
melt-formed surfaces can be described in a satisfactory manner by means
of frozen capillary waves, surfaces originating from a fracture process
exhibit a logarithmic growth of the height-height correlation, a result
that so far has not been obtained from microscopic calculations.  A recent
atomistic simulation study of metal-based materials (in both crystalline
and amorphous forms) has found that compression-induced rough surfaces are
self-affine on the length scale of 1-100~nm, a result that was attributed
to atomic-scale fluctuations in plastic flow~\cite{hinkle2020emergence}.
Together with our simulation results one thus can conclude that, for
amorphous solids, the surface topography on small length scales depends
strongly on the manufacturing history and the type of material considered.
Further research exploring how material composition and deformation mode
affects the surface topography will thus be very valuable.

We thank D. Bonamy for useful discussions.
Z.Z. acknowledges financial support by China Scholarship Council (NO. 201606050112).
W.K. is member of the Institut Universitaire de France.
This work was granted access to the HPC resources
of CINES under the allocation A0050907572 and A0070907572
attributed by GENCI (Grand Equipement National de Calcul
Intensif).

\normalem  

%

\clearpage
\newpage
\noindent{\Large \bf
Supplemental Material:\\ 
Roughness and scaling properties of oxide glass surfaces at the nanoscale}\\

\noindent Zhen Zhang$^1$, Simona Ispas$^1$, and Walter Kob$^1$\\
\noindent $^1${\it Laboratoire Charles Coulomb (L2C), University of Montpellier and CNRS, F-34095 Montpellier, France}\\[5mm]

In this Supplemental Material, we provide information regarding the
following points: 1) Details of the simulation and sample preparation, 2)
Details of the procedure used for constructing the surfaces, 3) Test
whether roughness depends on the distance to origin of the fracture,
4) Influence of strain rate and temperature on surface topography, 5)
Extracted parameters from the height-height correlation function, 6)
Details on the surface smoothing. \\

\renewcommand{\figurename}{Figure}
\renewcommand{\thefigure}{S\arabic{figure}}
\setcounter{figure}{0}

\noindent{\bf 1. Simulation details and sample preparation}\\[-6mm]

We consider glasses with the composition SiO$_2$ and Na$_2$O-$x$SiO$_2$,
NSx, with $x=3, 4, 5, 7, 10,$ and 20.  Approximately $2,300,000$
atoms were placed randomly in the simulation box which had a fixed
volume determined by the experimental value of glass density at
room temperature~\cite{bansal_handbook_1986}.  Using periodic boundary
conditions in three dimensions these samples were first melted and
equilibrated at 6000~K for 80~ps in the canonical ensemble ($NVT$)
and then cooled and equilibrated at a lower temperature $T_1$ (still
in liquid state) for another 160~ps.  The temperature $T_1$ ranges
from 3000~K for SiO$_2$ to 2000~K for NS3 (25 mole\% Na$_2$O), see
Ref.~\cite{zhang_surf-vib_2020,zhang_thesis_2020} for details.  Subsequently we cut the
sample orthogonal to the $z-$axis, and added an empty space, thus creating two free surfaces i.e.~the sample had the geometry of a slab. Periodic
boundary conditions were applied in all three directions. In order to
ensure that the two free surfaces do not interact with each other, the
thickness of the vacuum layer varied from 6~nm for silica to 14~nm for
NS3. These samples were then equilibrated at $T_1$ for 1.6~ns, a time
span that is sufficient to allow the reconstruction of the surfaces
and the equilibration of the interior of the samples. Subsequently the
liquid samples were cooled via a two-stage quenching: A cooling rate
of $\gamma_1=0.125$~K/ps was used to quench the samples from $T_1$ to
a temperature $T_2$ and a faster cooling rate $\gamma_2=0.375$~K/ps
to cool them from $T_2$ to 300~K. Finally, the samples were annealed
at 300~K for 800~ps. The temperature $T_2$ at which the cooling rate
changes was chosen to be at least 200~K below the simulation glass
transition temperature $T_g$, i.e.,~depending on 
composition, 1500~K $\geq T_2 \geq 800$~K, 
see Refs.~\cite{zhang_surf-vib_2020,zhang_thesis_2020} for details.  At $T_2$, we also
switched the simulation ensemble from $NVT$ to $NPT$ (at zero pressure)
so that the generated glass samples were not under macroscopic stress
at room temperature. 

After the glass samples were prepared we introduced on one of its
free surfaces a ``scratch'' in the form of a triangular notch spanning
the sample in the $y$-direction of width and depth of 3~nm and 2~nm,
respectively. Subsequently we applied to the sample a strain, with a
constant rate=0.5~ns$^{-1}$, until it broke. Due to the presence of the
notch, the place at which the fracture initiated could be changed at
will. More details can be found in Ref.~\cite{zhang_thesis_2020}.

The interaction between the atoms are given by a pairwise
effective potential proposed by Sundararaman et al. (SHIK)
~\cite{sundararaman_new_2018,sundararaman_new_2019}, which
has been found to give a good quantitative description
of the bulk and surface properties of sodo-silicate
glasses~\cite{zhang_potential_2020,zhang_surf-vib_2020}. Its functional
form is given by

\begin{equation}
V(r_{ij}) =  \frac{q_iq_je^2}{4\pi \epsilon_0 r_{ij}} +
A_{ij}e^{-r_{ij}/B_{ij}} - \frac{C_{ij}}{r_{ij}^6} \quad ,
\label{eq:potential}
\end{equation}

\noindent
where $r_{ij}$ is the distance between two atoms of species $i$ and $j$.
This potential uses partial charges $q_i$ for different atomic species:
The charges for Si and Na are, respectively, fixed at 1.7755$e$ and
0.5497$e$, while the charge of O depends on composition and is given by
requesting charge neutrality of the sample, i.e.,

\begin{equation}
q_{\rm O}=\frac{(1-y) q_{\rm Si}+2y
q_{\rm Na}}{2-y},
\end{equation}

\noindent
where $y$ is the molar concentration of Na$_2$O, i.e.,~$y=(1+x)^{-1}$. The other parameters
of the potential, $A_{ij}$, $B_{ij}$ and $C_{ij}$, occurring in
Eq.~(\ref{eq:potential}) are given in Refs.~\cite{sundararaman_new_2018,sundararaman_new_2019}.
Note that, following Ref.~\cite{sundararaman_new_2018}, the Coulombic
part in Eq.~(\ref{eq:potential}) was treated via the Wolf method.

Temperature and pressure were
controlled using a Nos\'e-Hoover thermostat and
barostat~\cite{nose_unified_1984,hoover_canonical_1985,hoover_constant-pressure_1986}.
All simulations were carried out using the Large-scale Atomic/Molecular
Massively Parallel Simulator software (LAMMPS)~\cite{plimpton_fast_1995} with a
time step of 1.6~fs.

The results presented in this manuscript correspond to one melt-quench sample
for each composition. However, we emphasize that the system sizes considered in this study are sufficiently large to make sample-to-sample fluctuations negligible. For the MS, the results for the two surfaces on the top and bottom sides of the glass sample were averaged. For the FS, twelve  surfaces, resulting from six independent fracture (by changing the location of the notch), were averaged. The error bars were estimated as the standard error of the mean of the samples. \\

\noindent{\bf 2. Identifying the surface}\\[-6mm]

In order to have a reliable description of the surface one needs a
method that allows to map the positions of the atoms onto a well-defined
mathematical surface.  The algorithm that we use for constructing this
surface mesh is based on the alpha-shape method of Edelsbrunner and
M\"ucke~\cite{edelsbrunner_three-dimensional_1994}. It starts with the
Delaunay tetrahedrization of the input point set, i.e. the atoms in
the sample.  From the resulting tetrahedra, all tessellation elements
are then evaluated by comparing their circumspheres to a reference probe
sphere that has a radius $R_\alpha$. The elements (with circumsphere
radius $R$) which satisfy $R < R_\alpha$ are classified as solid, and the
union of all solid Delaunay elements defines the geometric shape of the
atomistic solid. A robust realization of this algorithm is implemented
in OVITO ~\cite{stukowski_computational_2014}.

It is important to mention that the probe sphere radius $R_\alpha$ is
the length scale which determines how many details and small features
of the solid's geometric shape are resolved. To construct the geometric
surfaces for the glass samples, we use $R_\alpha = 3.2$~\AA\ , i.e.,
the typical distance between neighboring Si atoms. This choice allows
to resolve fine surface features and avoids artificial holes in the
constructed surfaces. We note, however, that a small change of $R_\alpha$
(e.g. $\pm0.5$~\AA) will not alter significantly the results presented in
the manuscript, see~Refs.~\cite{zhang_surf-vib_2020,zhang_thesis_2020}
for details. Finally, we mention that for the FS we have not used
for the analysis the parts of the surface that are closer than $\approx5$~nm to the
top/bottom MS in order to avoid the influence of these surfaces onto
the properties of the FS.

Once the geometric surface is constructed, i.e., the mesh points of
the surface are identified, we first fit a plane to the set of mesh points
using a least squares fitting procedure. Finally, a linear interpolation
is applied to the triangular mesh to obtain a uniform quadratic grid
which is subsequently used to determine the morphology and roughness of
the surface.\\

\clearpage
\noindent{\bf 3. Dependence of roughness on the distance from the fracture origin}\\[-6mm]

Since the formation of the FS is a strongly out-of-equilibirum
process it could be that the roughness depends on the distance
the crack has propagated from its origin, i.e., the notch. To test for this
possibility we have divided the surface into four segments along
the $y$-direction, i.e., in the direction the crack propagated,
see Fig.~\ref{SI_fig1_surface-segment-roughness}(a). For each
of these segments we have determined the roughness $\sigma$ and in
Fig.~\ref{SI_fig1_surface-segment-roughness}(b) we show for selected
glass compositions the value of $\sigma$ in these segments. The graph shows
that within the accuracy of the data there is no dependence on the segment
number, i.e., on the distance from the fracture origin.\\

\begin{figure}[ht]
\centering
\includegraphics[width=\columnwidth]{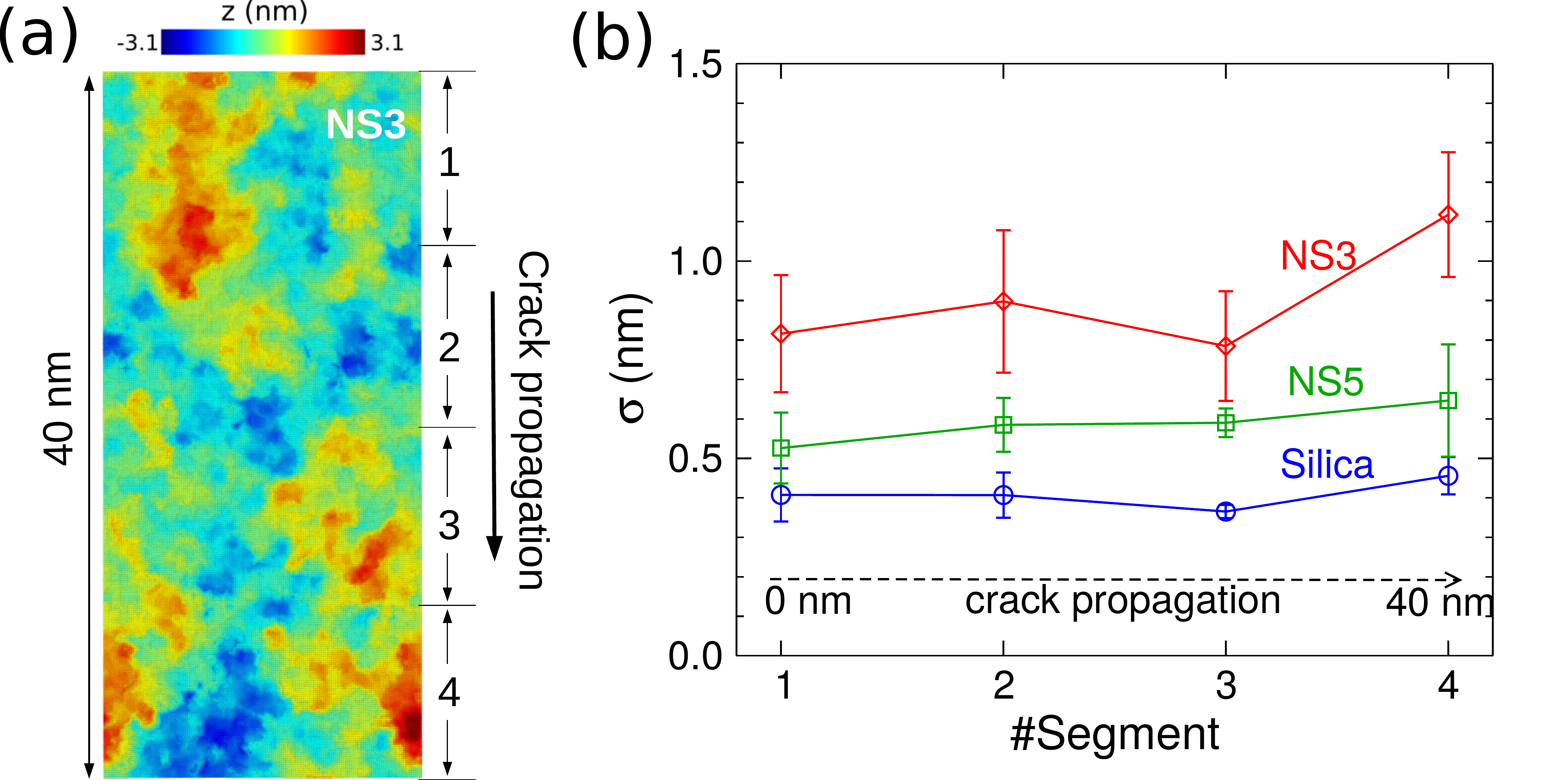}
\caption{(a) Height map $z(x,y)$ of a fracture surface of the NS3
glass. The surface is divided into 4 equal segments along the crack
propagation direction.  (b) Roughness $\sigma$ of the surface segments along
the crack propagation direction for three different glass compositions.  }
\label{SI_fig1_surface-segment-roughness}
\end{figure}

\noindent{\bf 4. Dependence of roughness and topography on strain rate and temperature}\\[-6mm]

The results in the main text have been obtained for a strain rate of
0.5~ns$^{-1}$.  Since it is well known that the properties of glasses
depend on the production history~\cite{binder_glassy_2011}, it can be expected that
also the properties of the fracture surface will depend on these details.
One key parameter for the fracture process is the strain rate, $\dot{\epsilon}$,
used to deform the sample and in Fig.~\ref{SI_fig2_ns0-FS-srate-effect}(a)
we show for the case of silica the roughness $\sigma$ as a function
of this parameter. One sees that for high rates the roughness
is about a factor of two higher than the $\sigma$ we obtain for
$\dot{\epsilon}=0.5$~ns$^{-1}$. This result is reasonable since a
high rate will not allow the crack to find energetically favorable
pathways and hence results in a surface that is rough. However,
once $\dot{\epsilon}$ is lowered to 0.5~ns$^{-1}$, $\sigma$ is
basically independent of the strain rate and hence we can conclude that
the roughness we present in the main text should correspond to the case
of real experiments on dynamic fracture.

A similar conclusion can be reached for the topography of the surface
since, see Fig.~\ref{SI_fig2_ns0-FS-srate-effect}(b), the height-height
correlation functions for the intermediate and low strain rates are basically identical.

\begin{figure}[bth]
\centering
\includegraphics[width=\columnwidth]{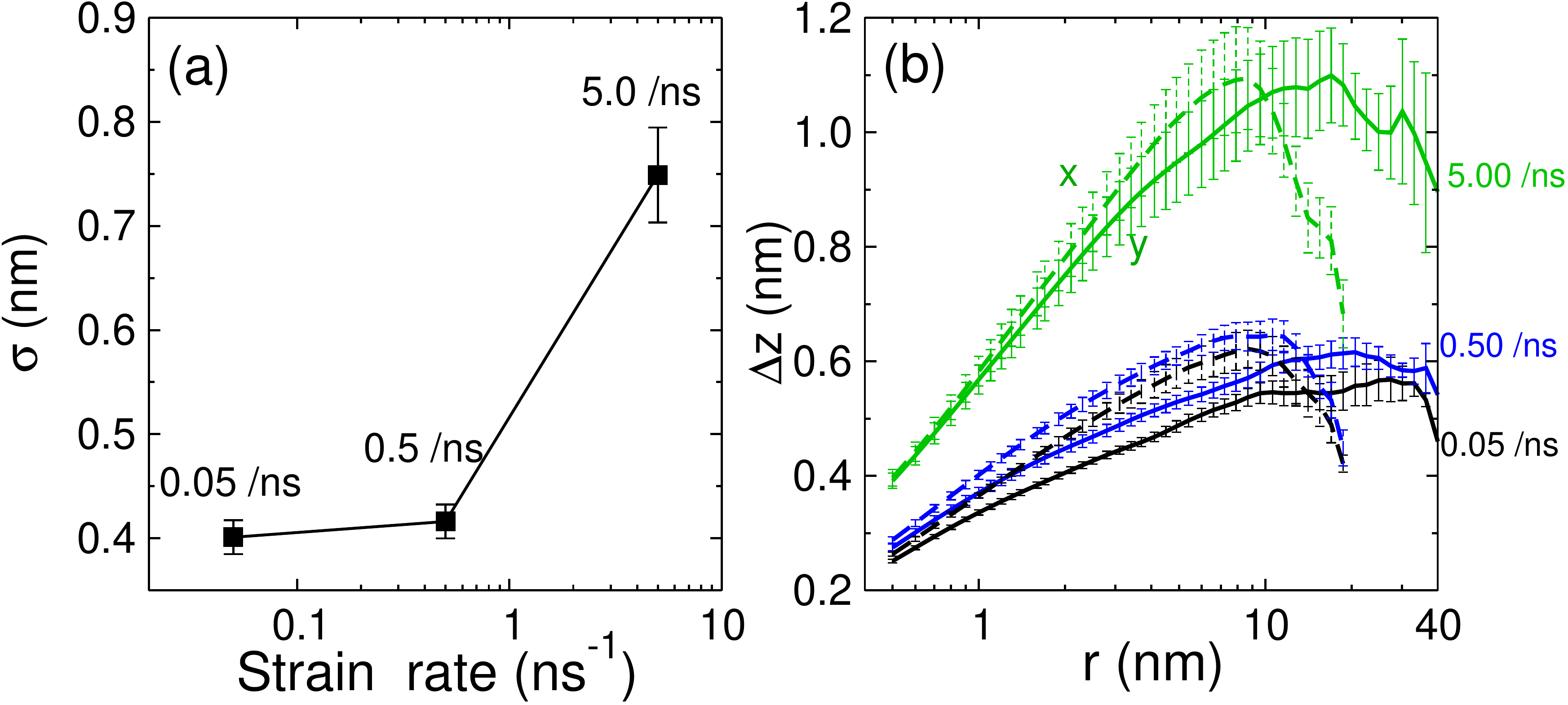}
\caption{Influence of strain rate on the roughness and scaling property
of the fracture surface for the case of silica. The data are plotted on
linear-log scale. In panel (b), $x$ and $y$ correspond to the direction
parallel to the crack front (dashed lines) and the direction of crack
growth (solid lines), respectively.}
\label{SI_fig2_ns0-FS-srate-effect}
\end{figure}

A further important parameter that influences the topography of a surface
is the temperature. The results presented in the main text are for the
temperature $T=300$~K. In Fig.~\ref{SI_fig3_nsx-surf-roughness-Tdepend}
we show how the roughness and the skewness $\gamma_1$ depend on the
temperature at which the surfaces are probed. To get this data for the
MS we used the configurations obtained during the quench procedure. For
the FS we annealed the samples at 600-800~K (NS3-silica) for 160~ps
before we cooled them down in a step-wise manner to the temperatures of
interest. At each $T$ we annealed the samples for 160~ps before starting
to strain them until fracture occurred.

The figure shows that for the case of the MS, panel (a), the roughness
of the silica surface is basically independent of $T$, i.e.~increasing
temperature does not allow the surface to fluctuate with significantly
larger amplitudes. This result is reasonable in view of the strong bonds
present in this kind of glass. In contrast to this we find for NS3
a  significant $T-$dependence in the roughness, although the absolute
change is small (10\%). 
This result is likely related to the fact that
the mobility of the Na atoms depends strongly on $T$, thus allowing for 
a significant softening of the network structure.	

\begin{figure}[th]
\centering
\includegraphics[width=0.9\columnwidth]{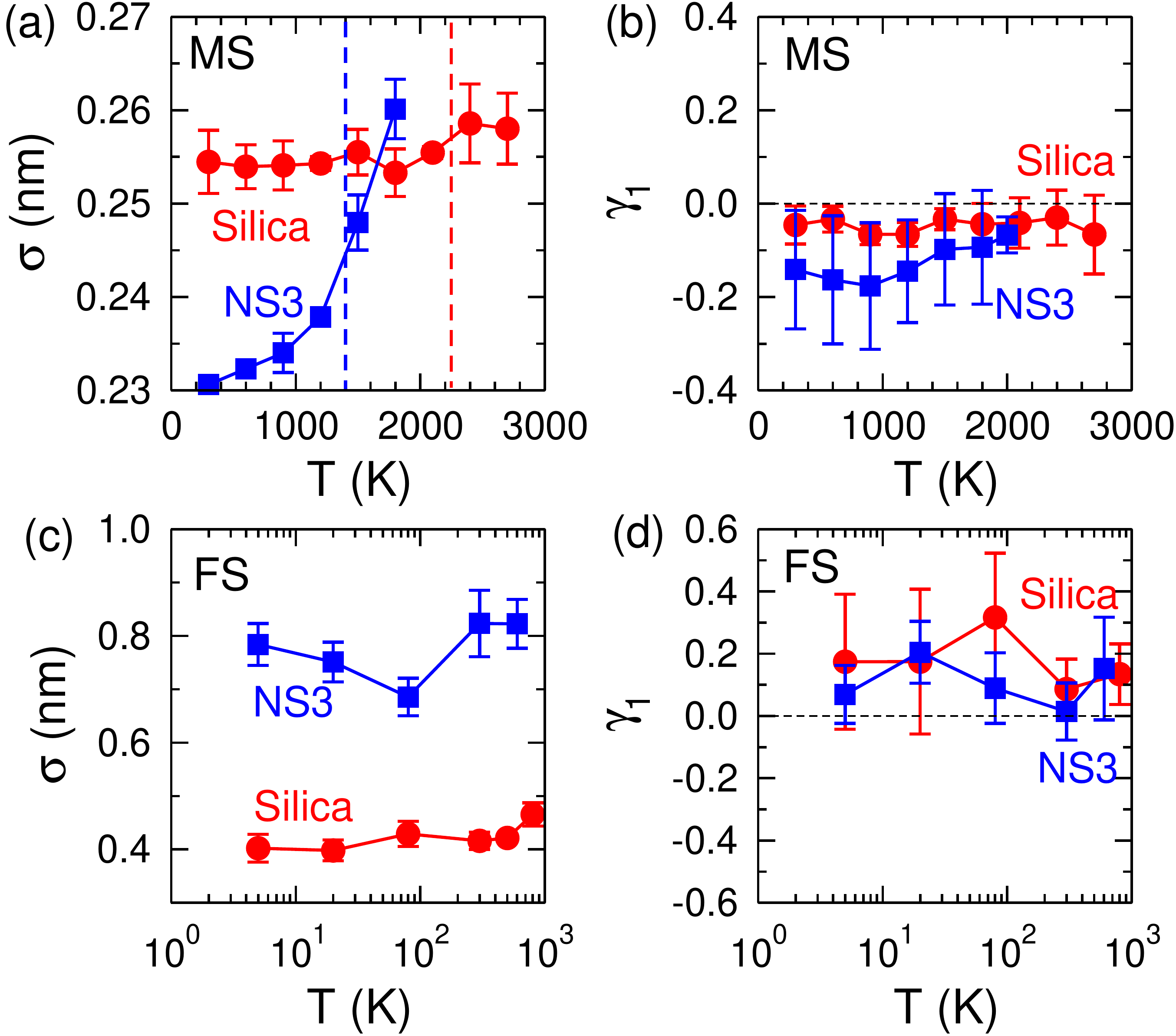}
\caption{Effect of temperature on surface properties. Upper and lower
panels are for the MS and FS, respectively. (a) and (c): $T-$dependence
of surface roughness. Vertical dashed lines in (a) indicates the $T_g$
of the silica and NS3 glasses. (b) and (d): The skewness characterizes the
asymmetric property of the surfaces.  }
\label{SI_fig3_nsx-surf-roughness-Tdepend}
\end{figure}

Also included in the graph are the estimated values for the glass
transition temperature $T_g$ (vertical dashed lines). Naively one
can expect that below $T_g$ the topography of the surface is given by
the thermal (harmonic) fluctuation of the atoms around their average
positions, i.e., a $T^{1/2}$ dependence. The NS3 data clearly shows
that this view is too simplistic since, as mentioned above, a change
of $T$ leads to a modification of the local elastic constants and as a
consequence the $T-$dependence of $\sigma$ is much stronger than expected
for a purely harmonic system.

Panel (b) shows that temperature does not really affect the skewness of
the distribution, i.e.~the fact that the surface has more deep cavities
than high peaks is independent of $T$. Only for the case of NS3 we find
at intermediate and high $T$ a slight increase of $\gamma_1$, i.e.,
at these high temperatures the system becomes sufficiently soft that
the distribution of the fluctuation becomes symmetric since the sample
starts to liquify.

For the case of the FS, panel (c), we see that neither silica nor NS3
show a significant $T$-dependence of $\sigma$. This observation is
coherent with the observation, Ref.~\cite{zhang_thesis_2020}, that during the fracture process the {\it local} temperature of the system close to the
crack front is so high (because of the breaking of the bonds) that for
the resulting surface it is irrelevant at which temperature the fracture
happens. In agreement with this argument we find that also the skewness
of the height distribution, panel (d), is independent of $T$.\\

\noindent{\bf 5. Parameters describing the height-height correlation function}\\[-6mm]

In the main text we have shown that the functional form describing the
height-height correlation function $\Delta z(r)$ depends on the type of
surface considered: $[\Delta z(r)]^2 = A\ln (r/\lambda)$ for the MS and $\Delta
z(r) = A \ln (r/\lambda)$ for the FS, see Fig.~\ref{fig3_surface-height-corr}.
In Fig.~\ref{SI_fig4_nsx-surf-hh-MS-FS-fitting} we show how the
prefactor $A$ and the length scale $\lambda$ depend on the composition
of the glass. Panel~(a) shows the prefactor $A$ of the logarithmic
dependence for the case of the MS. Although the scattering of the data is
substantial, there is good evidence that $A$ has a maximum at intermediate
concentrations of Na. One possibility to rationalize this observation
is to recall that within the frozen capillary wave theory this prefactor
is given by~\cite{sarlat_frozen_2006}

\begin{equation}
A=\frac{k_BT_f}{\pi \gamma} \quad,
\label{eq2}
\end{equation}

\noindent
where $\gamma$ is the surface tension. The freezing temperature $T_f$
decreases with increasing Na concentration, and it can be expected that
the surface tension does the same. Hence $\gamma^{-1}$ is an increasing
function of the Na$_2$O concentration and thus it is not unreasonable
(although not guaranteed) that the product of these two factors give a
maximum at intermediate values of the Na$_2$O content.

\begin{figure}[th]
\centering
\includegraphics[width=0.9\columnwidth]{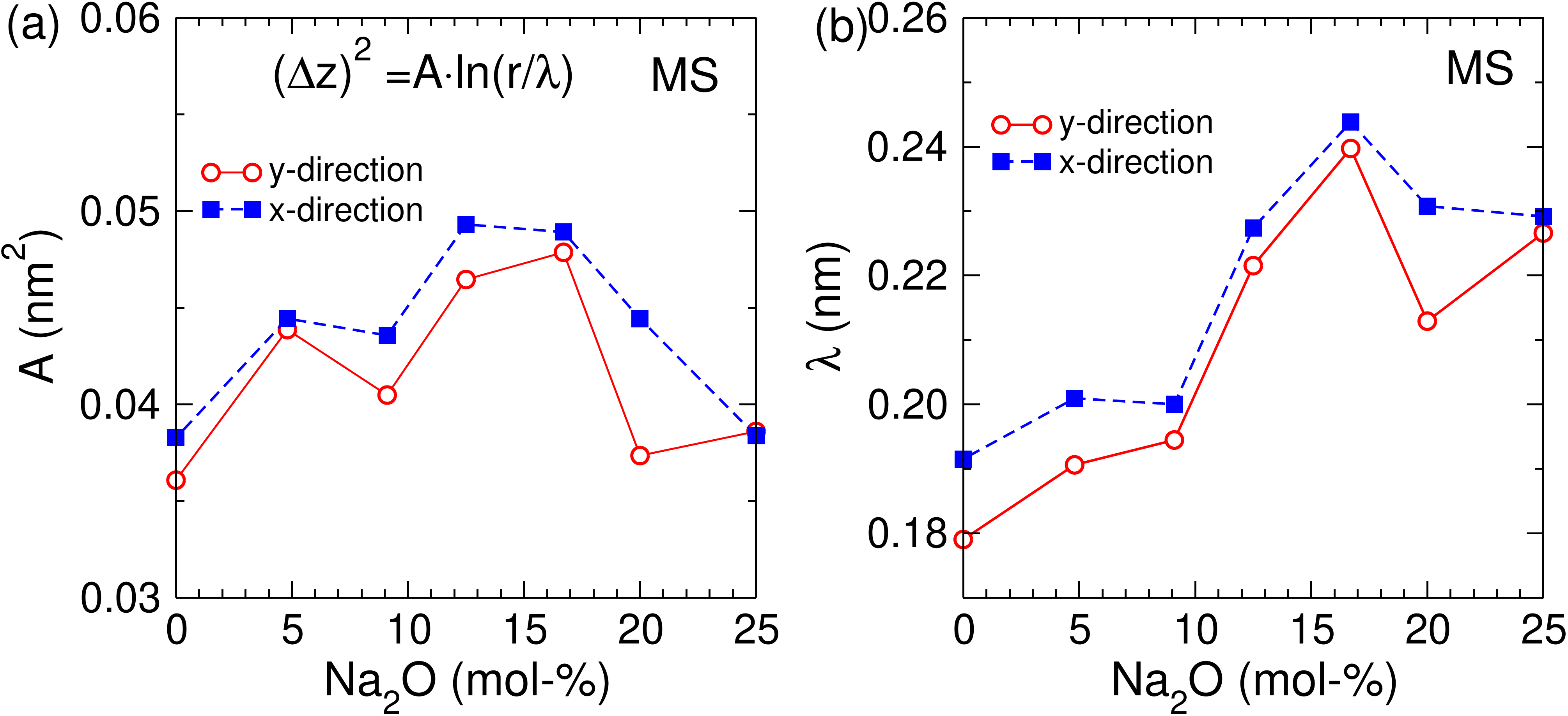}
\includegraphics[width=0.9\columnwidth]{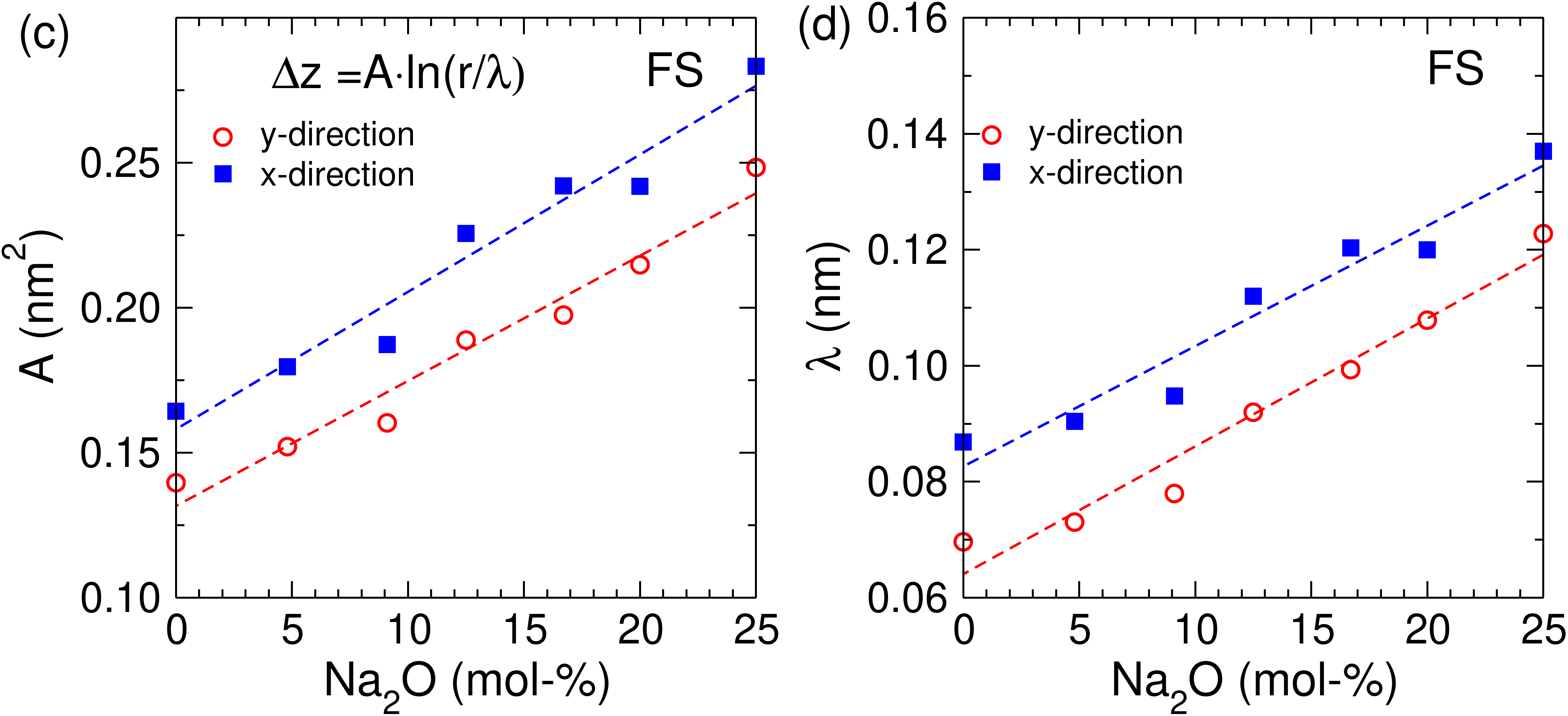}
\caption{Parameters of the logarithmic fit to the height correlation
function of the surfaces at 300~K. (a) and (b) are for the MS, and (c)
and (d) are for the FS. The fitting was performed only for the data at
$r<1$~nm, see Fig.~\ref{fig3_surface-height-corr}. The expressions used
for the fit are given in the graphs as well. $x$ and $y$ corresponds to
the direction parallel to the crack front and the direction of crack
growth, respectively. The lines in (b) and (d) are linear fits to the
data sets.
}
\label{SI_fig4_nsx-surf-hh-MS-FS-fitting}
\end{figure}

Panel (b) shows the compositional dependence of the length scale $\lambda$ and we see
that with increasing Na$_2$O the length scale increases weakly. This
observation agrees with the idea of the capillary wave theory which
identifies $\lambda$ as the smallest length scales over which one can use
this approach. Since it is well known that systems containing Na have
a more complex structure than pure silica, because of the presence of
ion-conducting channels~\cite{greaves_exafs_1985,horbach_dynamics_2002,meyer_channel_2004}, we can expect that
this minimum wave-length is larger for NS3 than for silica, rationalizing
the observed trend in panel~(b).

For the case of the FS, panels (c) and (d), we find that the prefactor $A$ as well as
the length scale $\lambda$ increase significantly (by more than 50\%) in
the $x$-range considered. Although we have no explanation why, for
both quantities, this dependence is linear in the Na concentration,
this qualitative trend is reasonable since an increasing Na content
will increase the plasticity of the system and hence allow for a
height-height correlation that has a larger amplitude and extents to
larger distances.\\

\noindent{\bf 6. Smoothing of the surface} \\[-6mm]

In the main text we have presented the results on the height-height
correlation function once the surface fluctuations were smoothed by
convoluting them with a 2D Gaussian. To do this smoothing of the surface
defined by points that are on the quadratic grid we proceeded as follows:
The weight function is given by

\begin{equation}
f(r)=\frac{1}{2\pi\omega^2}e^{-\frac{r^2}{2\omega^2}},
\end{equation}

\vspace*{2mm}
\noindent
where $r=\sqrt{(x-x_0)^2+(y-y_0)^2}$ is the in-plane distance from a
given grid point $(x,y)$ to the reference grid point $(x_0,y_0)$,
and $\omega$ controls the shape of the weight function. We consider
only the grid points with $r<2\omega$. Thus the smoothed surface height
$z'(x_0,y_0)$ is given by

\begin{equation}
z'(x_0,y_0)=\frac{\sum_{i=1}^N z(x_i,y_i)f(r_i)}{\sum_{i=1}^N f(r_i)},
\end{equation}

\noindent
where $N$ is the number of grid points that satisfy $r<2\omega$. By
varying the value of $\omega$, different levels of smoothing can be
applied to the original surface and thus the influence of spatial
resolution on the surface properties can be investigated.

\end{document}